\documentclass[pra,twocolumn,10pt,aps,showpacs,floatfix,amsmath,amsfonts,letterpaper,superscriptaddress]{revtex4-1}

\usepackage{stix}
\usepackage{hyperref}
\usepackage{graphicx}

\newcommand{\ee}{\mathrm{e}}
\newcommand{\ii}{\mathrm{i}}

\newcommand{\del}{\vec{\nabla}}
\newcommand{\Av}{\vec{A}}

\newcommand{\deltav}{\vec{\delta}}

\newcommand{\sigmamv}{\vec{\boldsymbol{\sigma}}}

\newcommand{\rv}{\vec{r}}

\newcommand{\Nv}{\vec{N}}

\newcommand{\zv}{\boldsymbol{z}}

\newcommand{\scL}{\mathcal{L}}
\newcommand{\scZ}{\mathcal{Z}}
\newcommand{\scC}{\mathcal{C}}
\newcommand{\scN}{\mathcal{N}}
\newcommand{\coord}{\mathfrak{z}}

\newcommand{\SO}{\mathrm{SO}}
\newcommand{\rmU}{\mathrm{U}}

\newcommand{\SONphi}{\SO(2)_{{\chi}}}

\makeatletter
\def\beq{\@ifstar{\@ifnextchar[{\@beqslabel}{\@beqsnolabel}}%
{\@ifnextchar[{\@beqlabel}{\@beqnolabel}}}
\def\@beqlabel[#1]{\begin{equation}\label{#1}}
\def\@beqnolabel{\begin{equation}}
\def\@beqslabel[#1]{\begin{equation*}\label{#1}}
\def\@beqsnolabel{\begin{equation*}}
\def\eeq{\@ifstar{\end{equation*}}{\end{equation}}}
\makeatother

\newcommand{\refeq}[1]{Eq.~(\ref{#1})}

\newcommand{\refcite}[1]{Ref.~\cite{#1}}

\newcommand{\reffig}[1]{Fig.~\ref{#1}}

\newcommand{\punc}[1]{\,{\text{#1}}}
\newcommand{\sub}[1]{_{\text{#1}}}

\newcommand{\spr}[1]{^{(#1)}}

\newcommand{\NCCP}{NCCP\(^1\)}

\newcommand{\degree}{^{\circ}}

\usepackage{color}

\begin{document}

\title{Emergent SO(5) symmetry at the columnar ordering transition in the classical cubic dimer model}


\author{G. J. Sreejith}
\affiliation{Indian Insitute for Science Education and Research, Pune 411008, India}

\author{Stephen Powell}
\affiliation{School of Physics and Astronomy, The University of Nottingham, Nottingham, NG7 2RD, United Kingdom}

\author{Adam Nahum}
\affiliation{Theoretical Physics, Oxford University, 1 Keble Road, Oxford OX1 3NP, United Kingdom}

\begin{abstract}
The classical cubic-lattice dimer model undergoes an unconventional transition between a columnar crystal and a dimer liquid,  in the same universality class as the deconfined quantum critical point in spin-1/2 antiferromagnets but with very different microscopic physics and microscopic symmetries.
Using Monte Carlo simulations, we show that this transition has an
emergent \(\SO(5)\) symmetry relating quantities characterizing the
two adjacent phases.
While the low-temperature phase has a conventional order parameter,
the defining property of the Coulomb liquid on the high-temperature
side is deconfinement of monomers, and so the \(\SO(5)\) symmetry
relates fundamentally different types of objects.
We study linear system sizes up to \(L=96\), and find that
this symmetry applies with an excellent precision that consistently
improves with system size over this range.
It is remarkable that SO(5) emerges  in a system as basic as the cubic dimer model, 
with only simple discrete degrees of freedom. 
Our results are important evidence for the generality of the $\SO(5)$ 
symmetry that has been proposed for the \NCCP\ field theory.
We describe a possible interpretation for these results in terms of a
consistent hypothesis for the renormalization-group flow structure, 
allowing for the possibility that $\SO(5)$ may ultimately be a near-symmetry rather than an exact one.
\end{abstract}

\maketitle

The classical dimer model on the cubic lattice illustrates three key mechanisms in three-dimensional (3D) critical phenomena. Two of these are the appearance of artificial gauge fields, and unconventional phase transitions at which topological effects play a key role. The third, which we demonstrate here, is the emergence in the infrared (IR) of unusual non-abelian symmetries that would be impossible at a conventional Wilson--Fisher-like critical point.

The close-packed dimer model has a power-law correlated `Coulomb'  phase \cite{Huse2003,Henley2010}, governed by an emergent \(\rmU(1)\) gauge field whose conserved flux arises from a microscopic `magnetic field' defined in terms of dimers. A remarkable phase transition \cite{Alet2006} separates this Coulomb liquid from a `columnar' phase, in which the dimers form a crystal that breaks lattice symmetries spontaneously. Despite being entirely classical, this transition is not described by Ginzburg--Landau theory, but is instead known to be a Higgs transition in the language of the $\mathrm{U}(1)$ gauge theory \cite{PowellChalker,Charrier2008,Chen2009}. The effective field theory is the noncompact $\mathrm{CP}^1$ model (\NCCP), in which the the gauge field is coupled to a two-component bosonic matter field that condenses at the transition.

\NCCP\ is also the effective field theory for the `deconfined' N\'eel---valence-bond solid (VBS) phase transition \cite{Senthil2004,Senthil2004a}  in 2+1D quantum antiferromagnets \cite{sandvikPRL2007,MelkoKaul,LouSandvikKawashima,BanerjeeDamleAlet,Sandvik2010,Harada2013,Jiang2008,Chen2013,Kuklov2008,ShaoGuoSandvik}
and  a related lattice loop model \cite{Nahum2015a}. This raises the possibility that the dimer model may exhibit a surprising emergent symmetry: Numerical simulations of the loop model for the N\'eel--VBS transition show \(\SO(5)\) symmetry emerging at large scales \cite{Nahum2015}---either exactly or to an extremely good approximation. 
Seminal earlier work on topological sigma models for deconfined critical points \cite{Tanaka2005,Senthil2006} had revealed that 
$\SO(5)$ is a consistent possibility in the IR, despite the fact that it cannot be made manifest in the gauge theory description \cite{anomalyfootnote}.
The N\'eel--VBS transition involves a three-component antiferromagnetic order parameter and a two-component VBS order parameter, and \(\SO(5)\) allows all five components of these order parameters to be rotated into each other.
This symmetry can be understood in terms of a set of dualities/self-dualities for \NCCP\ and related theories \cite{Wang2017}.

In this paper we use Monte Carlo simulations to demonstrate emergent \(\SO(5)\) at the dimer ordering transition. This large symmetry is particularly striking because the discrete classical model has no internal symmetries at all, only spatial symmetries together with a local constraint that is equivalent to a \(\rmU(1)\) symmetry in a dual representation. \(\SO(5)\) furthermore unifies operators of conceptually distinct types, rotating the crystal order parameter---a conventional observable in terms of the dimers ---into `monopole' operators that insert or remove monomers, and cannot be measured in the ensemble of dimer configurations. Together these yield a five-component \(\SO(5)\) superspin. The emergent symmetry group is therefore identical to that of the N\'eel--VBS transition, but it should be noted that the microscopic symmetries of the latter---roughly speaking, $\mathrm{SO}(3)\times \text{(lattice symmetries)}$---are very different from the $\text{(lattice symmetries)}\times \mathrm{U}(1)$ present in the dimer model.

Previously, \(\SO(5)\) has been demonstrated directly only in a single lattice model \cite{Nahum2015}, and is also supported by  level degeneracies in the JQ model \cite{Suwa2016}, both realizations of the N\'eel--VBS transition. Its presence in the dimer model is particularly important because the IR behaviour of the \NCCP\ model is subtle and remains controversial \cite{Sandvik2010,Harada2013,Jiang2008,Chen2013,Kuklov2008,Bartosch,ShaoGuoSandvik,SimmonsDuffin,Nakayama,Nahum2015a,Wang2017}. The simplest explanation for \(\SO(5)\) would be flow to a fixed point at which allowed \(\SO(5)\)-breaking perturbations are irrelevant, but there are reasons for doubting that this occurs
\cite{scalingfootnote}.
The transition may ultimately be first order, with an exceptionally large but finite correlation length.
However, even in the absence of a true continuous transition, the RG flows for \NCCP\ may ensure `quasiuniversality' relating this class of transitions  \cite{Nahum2015a,Wang2017}, and approximate \(\SO(5)\) symmetry.
In this scenario \(\SO(5)\) is an approximate symmetry rather than an exact one, but can hold to higher accuracy than standard finite-size scaling.

Given this complexity it is important to test the robustness of \(\SO(5)\).
Finding \(\SO(5)\) symmetry in the dimer model provides crucial confirmation that this is a generic property of the models that have been argued to coarse-grain to \NCCP, rather than one requiring further fine-tuning. The most striking feature of the results we present here is that at the lengthscales accessible numerically (up to linear size \(L=96\)), \(\SO(5)\) is indistinguishable from an exact symmetry of the IR theory.

\paragraph{Dimer model} We consider a classical statistical model where the degrees of freedom are dimers on the links of a cubic lattice. Defining \(d_\mu(\rv) \in \{0,1\}\) as the occupation number on the link joining site \(\rv\) to its neighbor \(\rv + \deltav_\mu\) (where \(\deltav_\mu\) is a unit vector), the number of dimers at site \(\rv\) is given by
\beq
n(\rv) = \sum_\mu [ d_\mu(\rv) + d_\mu(\rv - \deltav_\mu) ]\punc.
\eeq
Close-packed dimer configurations are those where \(n(\rv) = 1\) for all sites \(\rv\). For any function \(F\) of the dimer configuration, let \(\langle F \rangle = \scZ_0^{-1}\sum_{\psi\in\scC_0}F(\psi)\ee^{-E_\psi/T}\) be the average over the ensemble \(\scC_0\) of close-packed dimer configurations, where \(E_\psi\) is the energy of configuration \(\psi\), \(T\) is the temperature (\(k\sub{B}=1\)), and \(\scZ_0\) is the partition function.

An equal-weight ensemble of all close-packed dimer configurations (i.e., with \(E_\psi = 0\)) is a Coulomb liquid phase with no long-range order, described by an emergent noncompact \(\rmU(1)\) gauge theory \cite{Henley2010}. Sites where \(n(\rv) \neq 1\) have charge \(Q(\rv) = (-1)^{r_x + r_y + r_z}[n(\rv) - 1]\) in this description, and are hence referred to as containing `monopoles'. A pair of test monopoles of opposite charge (for example, a pair of empty sites on opposite sublattices) is deconfined in the liquid phase.

The model can be driven into a confining, ordered phase by applying an energy \(-v_2 \scN_2\), where \(\scN_2\) is the number of nearest-neighbor parallel dimers \cite{Alet2006}. For \(T \ll v_2\), the dimers form a crystal that maximizes \(\scN_2\) and breaks the spatial symmetries (we choose units so $v_2=1$). An order parameter for this phase is the `magnetization density' \(\Nv\), defined by
\beq[EqDefinem]
N_\mu = \frac{2}{L^3}\sum_{\rv} (-1)^{r_\mu} d_\mu(\rv)\punc.
\eeq
As the temperature is increased through a critical value \(T\sub{c}\), there is a direct transition into the dimer liquid \cite{Alet2006}.

In fact, we use a configuration energy \(E_\psi = -v_2 \scN_2 + v_4 \scN_4\), where \(\scN_4\) is the number of cubes of the lattice containing four parallel dimers \cite{Charrier2010}. For $v_4>0$ this is a frustrating interaction which has the effect of decreasing \(T\sub{c}\). More importantly, as \(v_4\) is varied, the order of the transition changes from clearly first order (at negative \(v_4\)) to apparently continuous (positive \(v_4\)), with an apparent tricritical point close to \(v_4 = 0\) \cite{Charrier2010}. This point introduces complications in the scaling analysis, which can be avoided by using large positive \(v_4\).

There is no local order parameter for the liquid phase, which is instead characterized by the deconfinement of monopoles \cite{Huse2003}. Defining operators \(\varphi(\rv)\) and \(\bar\varphi(\rv)\) that respectively decrease and increase the monopole charge at \(\rv\), the monopole distribution function \(G\sub{m}\) can be expressed as
\beq[EqDefineGq]
G\sub{m}(\rv_+, \rv_-) \equiv \frac{1}{\scZ_0}\sum_{\psi\in \scC(\rv_+,\rv_-)}\ee^{-E_\psi/T} = \langle \bar\varphi(\rv_+)\varphi(\rv_-) \rangle\punc{.}
\eeq
Here, \(\scC(\rv_+,\rv_-)\) denotes the ensemble of dimer configurations that are close-packed except for a pair of monopoles of charge \(\pm 1\) at sites \(\rv_\pm\). In the liquid phase, \(G\sub{m}\) has a nonzero limit as \(\lvert\rv_+-\rv_-\rvert\rightarrow\infty\), indicating that monopoles are deconfined. We also define global operators \(\varphi = L^{-3}\sum_{\rv} \varphi(\rv)\) and \(\bar\varphi\).

\paragraph{Continuum theory}

A continuum action that is believed to describe the transition \cite{PowellChalker,Charrier2008,Chen2009} involves a noncompact \(\mathrm{U}(1)\) gauge field \(\Av\), minimally coupled to a 2-component complex vector \(\zv\),
\beq[EqCriticalAction]
{\scL} = \frac{\kappa}{2} \lvert \del\times\Av \rvert^2 + \lvert(\del - \ii \Av) \zv\rvert^2 + s\lvert\zv\rvert^2 + u(\lvert\zv\rvert^2)^2\punc{,}
\eeq
where \(s\) should be tuned to its critical value and \(u\) and \(\kappa\) are positive constants. In terms of this theory, often referred to as \NCCP, the local magnetization is \(\vec N \sim \zv^\dagger \sigmamv \zv\) and the monopole operator \(\bar\varphi(\rv)\) creates a source of the `magnetic field' \(\del\times\Av\). This continuum theory also describes the N\'eel--VBS transition in spin-$\frac{1}{2}$ antiferromagnets --- where \(\zv^\dagger \sigmamv \zv\) is the local N\'eel vector and \(\varphi\) is the complex order parameter for the VBS phase \cite{Senthil2004,Senthil2004a} ---    and the `hedgehog-free' $\mathrm{O}(3)$ model \cite{MotrunichVishwanath}.

\paragraph{\(\SO(5)\) symmetry}

The claim of \(\SO(5)\) symmetry is that the critical point has an emergent symmetry under \(\SO(5)\) rotations of the five-component order parameter
\beq
\Phi = (N_x,N_y,N_z,c\varphi_x,c\varphi_y)\punc{,}
\eeq
where \(\varphi_x=\frac{1}{2}(\varphi+\bar\varphi)\), \(\varphi_y=\frac{1}{2\ii}(\varphi-\bar\varphi)\), and \(c\) is a constant.
Microscopically the dimer model permits only discrete rotation and reflection symmetries involving \(\Nv\) \cite{symmetriesfootnote}, as well as the \(\SO(2)\) subgroup of rotations of \((\varphi_x,\varphi_y)\). The latter is equivalent to the \(\rmU(1)\) symmetry under \(\varphi \rightarrow \ee^{\ii\alpha}\varphi\), \(\bar\varphi\rightarrow \ee^{-\ii\alpha}\bar\varphi\) associated with the requirement of overall monopole charge neutrality, and corresponds to the global \(\rmU(1)\) symmetry of a dual description. In addition to these microscopic symmetries, we will demonstrate \(\SO(3)\) of \(\Nv\) and, crucially, a rotation symmetry for the two-component vector \(\chi = (N_x,c \varphi_x)\), which we denote \(\SONphi\). Together, this is sufficient to demonstrate full \(\SO(5)\) symmetry.

Because \(\varphi\) cannot be expressed as a function of the dimer degrees of freedom, it is not possible to measure a probability distribution of \(\chi\). Instead we will consider the implications of the putative \(\SONphi\) symmetry for the moments \(\langle\chi_\mu\chi_\nu\dotsm\rangle\). Some of these are automatically satisfied because of the microscopic symmetries under \(N_x \rightarrow -N_x\) and under \(\varphi_x \rightarrow -\varphi_x\). The first nontrivial equality is
\beq[EqQuadratic]
\langle N_x^2 \rangle = c^2 \langle \varphi_x^2 \rangle\punc{,}
\eeq
which implies that the ratio \(\langle N_x^2 \rangle/\langle \varphi_x^2 \rangle\) should be independent of system size at the transition.

If we define the normalized quantities \(\tilde{N}_x = N_x/\langle N_x^2 \rangle^{1/2}\) and \(\tilde{\varphi}_x= \varphi_x/\langle \varphi_x^2 \rangle^{1/2}\) with unit variance, then at quartic order, \(\SONphi\) implies
\beq[EqQuartic]
\langle \tilde{N}_x^4 \rangle = \langle \tilde{\varphi}_x^4\rangle = 3\langle \tilde{N}_x^2 \tilde{\varphi}_x^2 \rangle\punc{.}
\eeq
With access to expectation values up to fourth order in \(N_x\) and \(\varphi_x\), we can also check the sixth-order result
\beq[EqSixthOrder]
\langle \tilde{N}_x^2 \tilde{\varphi}_x^4 \rangle = \langle \tilde{N}_x^4 \tilde{\varphi}_x^2 \rangle
\punc{.}
\eeq

\paragraph{Numerical algorithm} Expectation values containing the operators \(\varphi\) and \(\bar\varphi\) are explicitly given by
\beq[EqDefinevarphi]
\left\langle F \bar{\varphi}^{m/2}\varphi^{m'/2}\right\rangle =\delta_{mm'}\frac{[(m/2)!]^2}{L^{3m}\scZ_0} \sum_{\psi \in \scC_m} F(\psi)\,\ee^{-E_\psi/T}\punc,
\eeq
where \(\scC_m\) is the set of all configurations with \(\frac{m}{2}\) monopoles of each sign. In our simulations, the only allowed monopoles are empty sites, or ``monomers''.  (As a result, the sum defining the global variable $\varphi$ runs only over a single sublattice, and that for $\bar \varphi$ runs over the other sublattice.) Allowing overlapping dimers would make no essential difference.

Our procedure for calculating such expectation values is based upon the standard worm algorithm \cite{ProkofevSvistunov,Sreejith2014,Sandvik2006}, but with an additional update that allows the monopole number to change. At each iteration, starting in a configuration with \(\frac{m}{2}\) monopoles of each sign, we either construct a worm, which gives a new configuration with the same \(m\), or apply a step that may change the monopole number. If \(m=0\), this involves attempting to remove a dimer, leaving behind a pair of neighboring monopoles of opposite charge. If \(m=4\) (i.e., if there are two monopoles of each sign), we attempt to add a dimer, annihilating two monopoles. If \(m=2\), either of these moves may be attempted, with fixed relative probability. The location for the attempted move is chosen randomly and the update is accepted with the standard Metropolis probability for the resulting energy change.

This procedure effectively samples from an ensemble with partition function
\beq
\scZ\sub{eff} = \sum_{m\in\{0,2,4\}} f_m \sum_{\psi\in\scC_m} \ee^{-E_\psi/T}\punc,
\eeq
where the weights \(f_m\) can be calculated in terms of the probabilities used at each step (whose values are chosen to optimize the algorithm; see Supplemental Material). Comparison with \refeq{EqDefinevarphi} then gives the desired quantities in terms of expectation values in this ensemble, conditioned on the number of monopoles.

\begin{figure}
\includegraphics[width=\columnwidth]{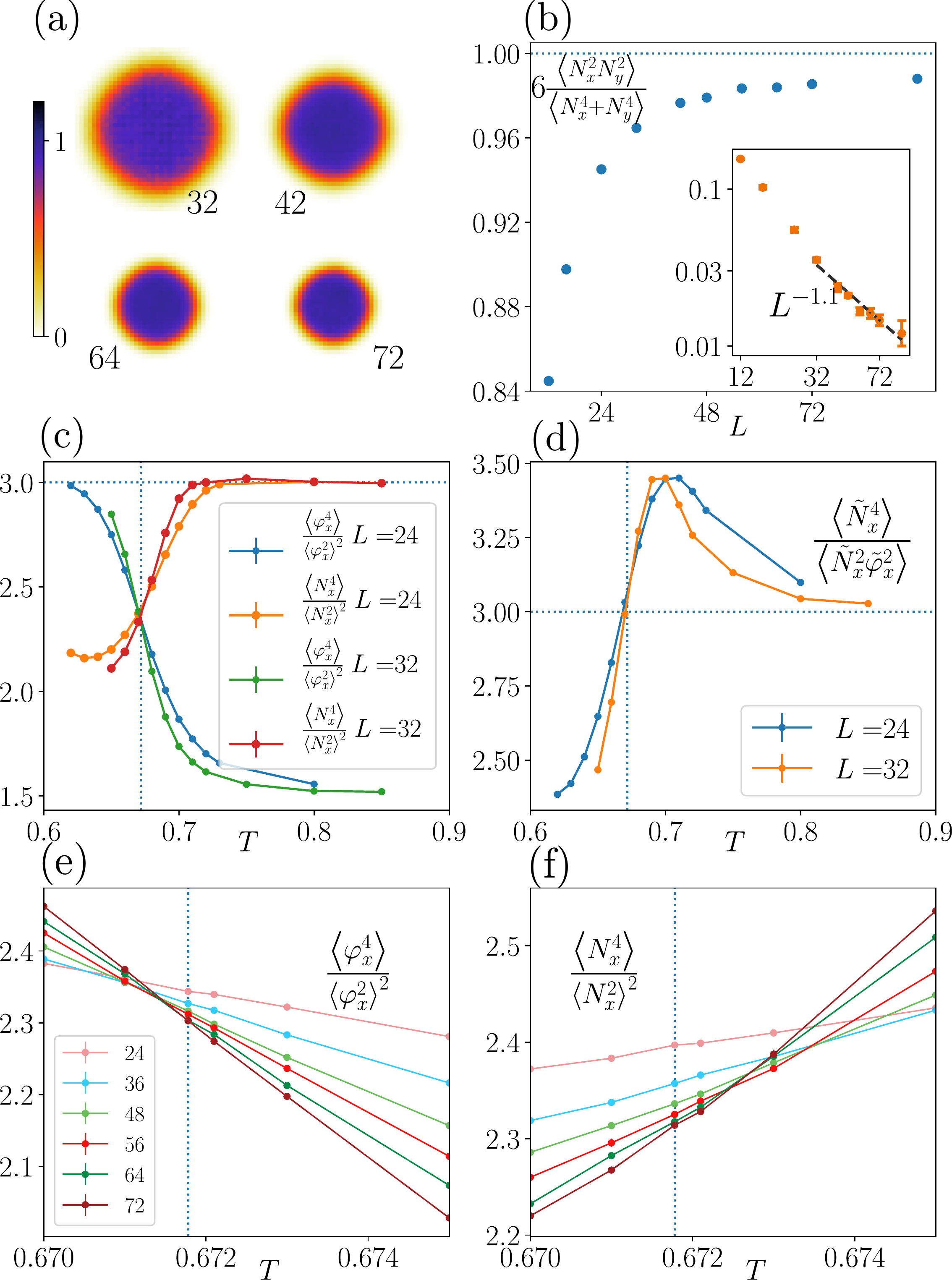}
\caption{Monte Carlo results across the columnar ordering transition for \(v_4 = 10\). (a) Cross section, with \(N_z =0\), through the magnetization density distribution at the critical temperature \(T\sub{c}\) [determined in \reffig{v4_10_plot2}(a)], labeled by system size \(L\). (b) Ratio of moments \(6\langle N_x^2 N_y^2 \rangle/\langle N_x^4 + N_y^4\rangle\), which is equal to unity in the case of \(\SO(3)\) symmetry, plotted as a function of system size \(L\) at the critical temperature \(T = T\sub{c}\). The inset shows the absolute difference between this ratio and unity, on a double-logarithmic scale, along with a fit to a power law. (c) Binder cumulants of the magnetization and of the monopole operator, for a broad range of temperatures. The dotted vertical line (in this and subsequent panels) shows the critical temperature \(T\sub{c}\). (d) Normalized cumulant ratio \(\langle \tilde{N}_x^4\rangle/\langle \tilde{N}_x^2 \tilde{\varphi}_x^2 \rangle\), which takes the value \(3\) in the case of \(\SO(5)\) symmetry. (e,f) Binder cumulants close to the critical temperature.}
\label{v4_10_plot1}
\end{figure}

\paragraph{Results} We first verify \(\SO(3)\) symmetry of \(\Nv\) at the critical point for \(v_4 = 10\), extending the results of \refcite{Misguich2008} at \(v_4 = 0\). Figure~\ref{v4_10_plot1}(a) shows a cross section (with \(N_z = 0\) \cite{pixelfootnote}) through the probability distribution for the magnetization \(\Nv\) at the critical temperature \(T\sub{c}\). (The value of \(T\sub{c}\) is determined using the procedure described below.) The circular distribution indicates that the microscopic symmetry under \(90\degree\) rotations of \(\Nv\) is enhanced to an emergent continuous symmetry at the critical point. A quantitative measure of this emergent symmetry is provided by the ratio \(6\langle N_x^2 N_y^2 \rangle/\langle N_x^4 + N_y^4\rangle\), which is plotted versus system size \(L\) at \(T\sub{c}\) in \reffig{v4_10_plot1}(b). This quantity approaches unity as the system size increases, indicating that the critical point has an emergent symmetry, at least under \(45\degree\) rotations. We argue that this provides strong evidence of emergent continuous \(\SO(3)\) symmetry.

We now turn to quantities that test for symmetry mixing \(\Nv\) and \(\varphi\), focusing on \(v_4 = 10\). In this case, distribution functions are not accessible, because \(\varphi\) cannot be expressed as an observable in the dimer ensemble \cite{jointdistributionfootnote}, and we therefore consider moments of the quantities \(N_x\) and \(\varphi_x\).

The Binder cumulants for the magnetization, \(\langle N_x^4 \rangle / \langle N_x^2 \rangle^2\), and for the monomer operator, \(\langle \varphi_x^4 \rangle / \langle \varphi_x^2 \rangle^2\), are shown in \reffig{v4_10_plot1}(c) for a broad range of \(T\). Both take the expected values deep within the two phases (see Supplemental Material) and they cross at approximately the same temperature, consistent with a continuous phase transition directly between the dimer crystal and dimer liquid. The first hint of \(\SO(5)\) symmetry is that the two Binder cumulants take the same value at their crossing points. It should be noted that this value is not what one would expect for a Gaussian probability distribution ({\it viz} \(3\)), which would provide a trivial explanation for our \(\SO(5)\) results. Next, Figure~\ref{v4_10_plot1}(d) shows \(\langle \tilde{N}_x^4\rangle/\langle \tilde{N}_x^2 \tilde{\varphi}_x^2 \rangle\) for the same range of temperatures. This is an example of a ratio that is constrained by \(\SO(5)\) [\refeq{EqQuartic}]. The ratio indeed takes the expected value of 3  at $T_c$, as we show to much higher precision below, while approaching the expected limits in both phases.

The above Binder cumulants for $N_x$ and $\varphi_x$ are shown in the neighborhood of their crossing points in \reffig{v4_10_plot1}(e,f). As has been noted in previous studies of this transition \cite{Charrier2010,Sreejith2014}, the crossings drift significantly with increasing \(L\), and the respective crossing temperatures for the two Binder cumulants differ by a relative amount of order \(10^{-3}\). 

\begin{figure}
\includegraphics[width=\columnwidth]{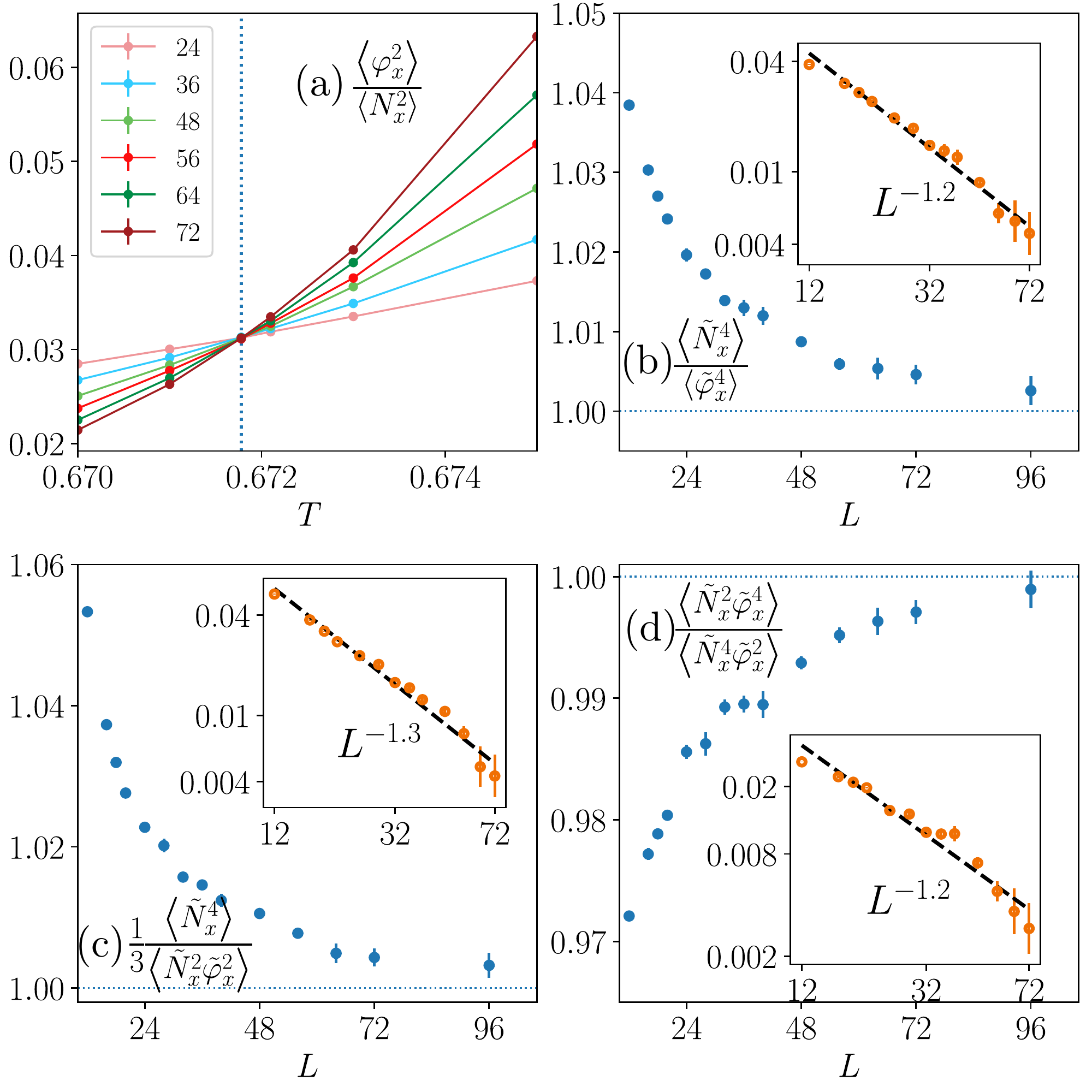}
\caption{Monte Carlo results at the columnar ordering transition for \(v_4 = 10\), demonstrating \(\SO(5)\) symmetry. (a) Ratio \(\langle \varphi_x^2 \rangle / \langle N_x^2 \rangle\) of second moments. The temperature of the crossing, shown with a vertical dashed line, is used as the value of \(T\sub{c}\) in the subsequent panels. (b--d) Ratios of cumulants, each of which is constrained to unity by \(\SO(5)\) symmetry. In each panel, the inset shows a log--log plot of the absolute difference of the ratio from unity versus \(L\), with a best-fit power law.}
\label{v4_10_plot2}
\end{figure}

However, the quantity \(\langle \varphi_x^2 \rangle / \langle N_x^2 \rangle\), which we show in \reffig{v4_10_plot2}(a), has a remarkably sharp crossing at a temperature in between the two. The presence of such a crossing is a consequence of \(\SO(5)\) symmetry [see \refeq{EqQuadratic}].
Since this crossing point is much better-defined than that of the Binder cumulants, we use it as our best estimate of the critical temperature, giving \(T\sub{c} = 0.6718(1)\) for \(v_4 = 10\).

\begin{figure}[t]
\includegraphics[width=\columnwidth]{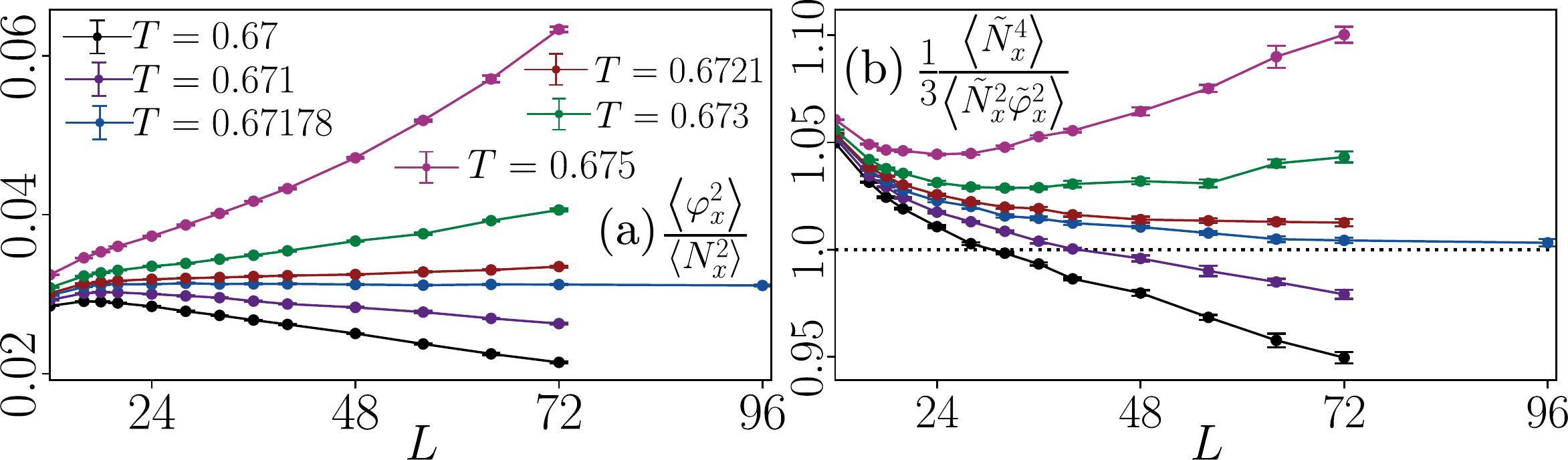}
\caption{{
Flow of moment ratios with linear system size $L$ for temperatures close to $T_c$.
(a) At estimate of $T_c$ (longer blue lines), $\langle \varphi_x^2\rangle / \langle N_x^2\rangle$ is independent of $L$ at large $L$ to a very good approximation; and (b)
$\frac{1}{3}\langle \tilde N_x^4\rangle / \langle \tilde N_x^2 \tilde \varphi_x^2 \rangle$ approaches unity. Both results are consistent with $\mathrm{SO}(5)$.
}}
\label{Fig3}
\end{figure}
The moment ratios \(\langle \tilde{N}_x^4\rangle/\langle \tilde{\varphi}_x^4 \rangle\), \(\frac{1}{3}\langle \tilde{N}_x^4\rangle/\langle \tilde{N}_x^2 \tilde{\varphi}_x^2 \rangle\), and \(\langle \tilde{N}_x^2 \tilde{\varphi}_x^4\rangle/\langle \tilde{N}_x^4 \tilde{\varphi}_x^2 \rangle\), evaluated at this temperature, are shown in \reffig{v4_10_plot2}(b--d). All three tend towards unity, 
as expected from
 \(\SO(5)\)  (Eqs.~\ref{EqQuartic} and~\ref{EqSixthOrder}).
The corrections are small in magnitude and decrease approximately as power laws in \(L\) $(\sim L^{-|y_\text{irr}|}$) over the full range of sizes studied. The  effective irrelevant exponent $y_\text{irr}$  is consistent with that observed for the corrections to the \(\SO(3)\) symmetry of \(\Nv\) [see \reffig{v4_10_plot1}(b), inset].

The same data for \(\langle \varphi_x^2 \rangle / \langle N_x^2 \rangle\) and \(\frac{1}{3}\langle \tilde{N}_x^4\rangle/\langle \tilde{N}_x^2 \tilde{\varphi}_x^2 \rangle\) is shown in \reffig{Fig3} as a function of \(L\) for various temperatures above and below the transition. As expected for an emergent symmetry, these ratios diverge from their critical values with increasing \(L\) at fixed temperatures on each side of \(T\sub{c}\).

In the Supplemental Material we show results for the transition at smaller values of  \(v_4\).  (Recall that there is a critical line in the plane of the frustrating interaction $v_4$ and temperature, with $v_4\simeq 0$ previously identified as the point separating continuous and first-order transitions \cite{Charrier2010}.) Importantly, results for $v_4 = 1$ are similar to $v_4=10$, with the ratios approaching their $\SO(5)$-invariant values to a similar level of precision at the largest sizes. 
Results at $v_4=0.2$ are still consistent with emergent $\SO(5)$, while at $v_4=0$, close to the apparent tricritical point, the deviation from unity is considerably larger at the largest sizes. Even here it is possible that $\SO(5)$ may improve at still larger sizes, but see discussion for another explanation.

We expect that the corrections to $\SO(5)$ arise from perturbations to a hypothetical $\SO(5)$-invariant continuum action, with these perturbations being effectively irrelevant at least on the scales we access. (If the critical properties are only `quasiuniversal', this $\SO(5)$-invariant action is not that of an RG fixed point, but is instead associated with a relatively well-defined $\SO(5)$-invariant flow line in coupling constant space.) 
Classifying operators into $\SO(5)$ representations, the simplest possibility is that the leading perturbations arise from an operator $X_{abcd}$ in a four-index tensor representation of $\SO(5)$ \cite{Nahum2015,Wang2017, operatorsfootnote}. The microscopic symmetries of the dimer model allow the perturbation $\sum_{a,b=1}^3 X_{aa bb}$, which is a higher-order asymmetry between $\varphi$ and $N$, and $\sum_{a=1}^3 X_{aaaa}$, which is cubic anisotropy for $N$.
The consistency of the $y_\text{irr}$ estimates  (in the range $-1.3$ to $-1.1$)  for distinct ratios at $v_4=10$ is in line with this picture. For $v_4=1$ we obtain slightly larger values (in the range $-1.7$ to $-1.4$), while in the loop model a rough estimate gave $y_\text{irr} \sim -0.8$. In the `quasiuniversal' picture $y_\text{irr}$ is an effective exponent which is expected to drift as a function of $L/L_0$, where $L_0$ is a nonuniversal lengthscale, so the differences in $y_\text{irr}$ may be attributable to such drifts.

\paragraph{Discussion and conclusions} We have presented evidence for \(\SO(5)\) symmetry at the ordering transition in the cubic dimer model, which is at least as robust as critical scaling. This is a remarkable example of an emergent symmetry in a purely classical model, relating the magnetization \(\Nv\) to the monopole operator \(\varphi\).

Because \(\varphi\) is not a local observable in terms of the dimer degrees of freedom, it is not possible to measure its distribution function. We have instead used the comoments of \(\varphi\) and \(\Nv\) to confirm invariance under a discrete subgroup of \(\SO(5)\) relating $\vec N$ and $\varphi$, as well as under the \(\SO(3)\) symmetry of \(\Nv\). This implies full \(\SO(5)\) symmetry.

These results demonstrate that very precise $\SO(5)$ is a robust property of a large class of models described by \NCCP. 
Still, it is possible that \NCCP\ does not have a true critical point, but instead has only `quasiuniversal' properties which are due to a near-vanishing of the beta function 
\cite{Nahum2015a,Wang2017}. 
The `critical' properties are then associated with an approximate convergence of the flows to an $\SO(5)$--invariant flow line, and  \(\SO(5)\) is an approximate symmetry rather than an exact one \cite{pseudocriticalfootnote}.
Our results are compatible with this scenario, although over the scales we have studied $\SO(5)$ resembles an exact symmetry of the IR theory, whose precision is improving with $L$. 

This scenario  allows for a sharp crossover, as a function of a microscopic coupling, between  a regime in which the transition is apparently continuous (very weakly first order) and one where it is strongly first order. The `tricritical' point $v_4\simeq 0$ might in fact be such a crossover. If so, a more detailed analysis of the \(\SO(5)\) ratios for $v_4\sim 0$ could give useful insight into the RG flows.
Another  extension would be to reduce the lattice symmetry of the model so that only 4 out of the 6  ordered states remain \cite{Chen2009}. This would yield a platform for investigating the possibility of emergent $\mathrm{O}(4)$ symmetry in the `easy-plane' version of \NCCP\ \cite{Wang2017,Meng2017,ZhangHe2018}.

\paragraph{Acknowledgments} We thank J.~T.~Chalker for useful discussions. This work was supported by EPSRC Grant No.\ EP/M019691/1 (SP) and Grant No. EP/N028678/1 (AN). We thank MPI-PKS Dresden and NPSF-CDAC Pune for providing computing resources. 

\clearpage

\appendix

\onecolumngrid
\section*{Supplemental material}

\setcounter{figure}{0}
\renewcommand\thefigure{S.\arabic{figure}}

\section{Numerical algorithm}

Suppose we have an algorithm that samples configurations at fixed monomer number. At each iteration, apply this step with probability \(1-p_m\) (\(p_m \ll 1\)), where \(m\) is the total number of monomers.

With probability \(p_m\), attempt to change the monomer number, carrying out the following attempted update with Metropolis acceptance rate (ignoring the monomer energy):
\begin{itemize}
\item If there are no monomers, pick a random dimer and attempt to remove it.
\item If there are two monomers, then:
\begin{itemize}
\item With probability \(q\) (\(q \simeq \frac{1}{2}\)), add a dimer if possible (i.e. if the monomers are adjacent); if not, do nothing.
\item With probability \(1-q\), pick a random dimer and remove it.
\end{itemize}
\item If there are four monomers, pick one of the monomers randomly and pick one of its links randomly. If a dimer can be added on the link do so; if not, do nothing.
\end{itemize}

To determine the occupation probabilities in the steady state, we check detailed balance. The probability of transition between any pair of configurations with the same monomer number \(m\) is multiplied by \(1-p_m\), and so detailed balance is unaffected.

Starting from a configuration \(\psi\) with zero monomers, the probability of transition to a configuration \(\psi'\), equal to \(\psi\) but with one  dimer removed, is \(\dfrac{2p_0}{L^3}R(\Delta E)\), where \(\frac{1}{2}L^3\) is the number of dimers and \(R(\Delta E)\) is the Metropolis acceptance probability associated with the energy change \(\Delta E\) (ignoring the monomer energy). The probability of the reverse process, a transition from \(\psi'\) to \(\psi\), is \(p_2 q R(-\Delta E)\). In the steady-state, the relative occupation probability of the two configurations is then
\beq
\frac{P(\psi)}{P(\psi')} = \frac{p_2 q R(-\Delta E)}{\tfrac{2p_0}{L^3}R(\Delta E)} = \frac{p_2}{p_0}\frac{q L^3}{2}\ee^{\Delta E /T}\punc.
\eeq
This is constant apart from the Boltzmann factor, and so determines the relative occupation of configurations in these two sectors.

Starting from a configuration \(\psi\) with two monomers, the probability of transition to a configuration \(\psi'\), equal to \(\psi\) but with  one dimer (on link \(\ell\)) removed, is \(\dfrac{p_2(1-q)}{\frac{1}{2}L^3 - 1}R(\Delta E)\). For the reverse probability, we need either of the two monomers on link \(\ell\) to be the one chosen (which happens with probability \(\frac{2}{4}\)) and then that link to be chosen (probability \(\frac{1}{\coord}\), where \(\coord\) is the coordination number), and so the probability is \(\dfrac{p_4}{2\coord}R(-\Delta E)\). The ratio of steady-state occupation probabilities is then
\beq
\frac{P(\psi)}{P(\psi')} = \frac{\dfrac{p_4}{2\coord}R(-\Delta E)}{\frac{p_2(1-q)}{\frac{1}{2}L^3 - 1}R(\Delta E)} = \frac{p_4}{p_2}\frac{\frac{1}{2}L^3 - 1}{2\coord (1-q)}\ee^{\Delta E / T}\punc.
\eeq
This is again constant apart from the Boltzmann factor.

The steady-state occupation probability for configuration \(\psi\) is therefore \(P(\psi) \propto f_{m(\psi)}\ee^{-E_\psi/T}\), where the overall constant is fixed by normalization and
\beq[EqSteadyState]
\begin{aligned}
f_0 &= 1\\
f_2 &= \dfrac{2}{qL^3}\dfrac{p_0}{p_2}\\
f_4 &= \dfrac{4\coord (1-q)}{(\frac{1}{2}L^3 - 1)qL^3}\dfrac{p_0}{p_4}\\
f_m &= 0 \text{ for \(m > 4\).}
\end{aligned}
\eeq
The parameters \(p_m\) (\(m \in \{0,2,4\}\)) and \(q\) can be chosen to optimize the algorithm. We are effectively sampling from an ensemble with partition function
\beq
\scZ\sub{eff}(\{f_m\}) = \sum_{m=0,2,4,\dotsc} f_m \sum_{\psi\in\scC_m} \ee^{-E_\psi/T}\punc.
\eeq

We can then calculate, for example, \(\langle N_x^2 \varphi_x^2\rangle\). To calculate this, start with \(Q\spr{0,0} = Q\spr{2,2} = 0\), and after every step of the algorithm, when the configuration is \(\psi\), change \(Q\spr{0,0}\) by
\beq
\Delta Q\spr{0,0}_\psi = \begin{cases}
1 & \text{if \(\psi\) has no monomers}\\
0 & \text{otherwise.}
\end{cases}
\eeq
and change \(Q\spr{2,2}\) by
\beq
\Delta Q\spr{2,2}_\psi = \begin{cases}
[N_x(\psi)]^2 & \text{if \(\psi\) has two monomers}\\
0 & \text{otherwise,}
\end{cases}
\eeq
In the steady state, their ratio is
\begin{align}
\frac{Q\spr{2,2}}{Q\spr{0,0}} &= \frac{\sum_\psi P(\psi) \Delta Q\spr{2,2}_\psi}{\sum_\psi P(\psi) \Delta Q\spr{0,0}_\psi}\\
&= f_2\frac{\sum_{\psi\in\scC_2} \ee^{-E_\psi/T}[N_x(\psi)]^2}{\sum_{\psi\in\scC_0} \ee^{-E_\psi/T}}\punc,
\end{align}
using \refeq{EqSteadyState} for the steady-state occupation probabilities \(P\), and so
\beq
\langle N_x^2 \varphi_x^2\rangle = \frac{q}{4L^3}\frac{p_2}{p_0}\frac{Q\spr{2,2}}{Q\spr{0,0}}\punc.
\eeq

More generally, change \(Q\spr{n,m}\) by
\beq
\Delta Q\spr{n,m}_\psi = \begin{cases}
[N_x(\psi)]^n & \text{if \(\psi\) has \(m\) monomers}\\
0 & \text{otherwise,}
\end{cases}
\eeq
and then the expectation value \(\langle N_x^n \varphi_x^m \rangle\) can be expressed in terms of the steady-state ratio \(\dfrac{Q\spr{n,m}}{Q\spr{0,0}}\).

\section{Moment ratios in phases}

In each phase one of $\vec N$, $\vec \varphi$ is disordered and the other is ordered. The disordered order parameter  has a Gaussian probability distribution (as it is the sum of almost independent contributions from different correlation volumes) and is decoupled from the ordered one. The ordered quantity should be treated as of fixed length and averaged over the manifold or set of symmetry-equivalent states.

{\bf Coulomb phase.} $\vec N$ is disordered so $\langle \tilde N_x^4 \rangle=3\langle\tilde N_x^2\rangle =3$, $\langle \tilde N_x^6\rangle=15$.  $ \tilde\varphi$ is averaged over the circle $(\tilde\varphi_x, \tilde\varphi_y) = \sqrt{2} (\cos \theta, \sin \theta)$ so that $\langle\tilde\varphi_x^2\rangle = 1$, $\langle\tilde\varphi_x^4\rangle= 3/2$:
\begin{align}
\langle \tilde N_x^4\rangle/\langle\tilde \varphi_x^4 \rangle &= 2&
\langle \tilde N_x^4 \rangle /\langle \tilde N_x^2 \tilde \varphi_x^2 \rangle &= 3 &
\langle \tilde \varphi_x^4 \rangle /\langle \tilde N_x^2 \tilde \varphi_x^2 \rangle &= 3/2.
\end{align}

{\bf Dimer ordered phase.} The ordered phase has two regimes because of the dangerous irrelevance of sixfold symmetry breaking at the critical point. In both regimes $\varphi$ is disordered: $\langle \tilde \varphi_x^4 \rangle = 3$.

{\bf $\mathrm{O}(3)$ symmetric regime (intermediate $T_c-T$).} We average $\vec N$ over the sphere $\vec{\tilde  N} = \sqrt{3} (\cos\theta, \sin \theta \cos \chi, \sin \theta \sin \chi)$ giving $\langle \tilde N_x^2\rangle = 1$, $\langle \tilde N_x^4 \rangle = 9/5$, $\langle \tilde N_x^6 \rangle =27/7$:
\begin{align}
\langle \tilde N_x^4\rangle/\langle \tilde \varphi_x^4 \rangle &=3/5&
\langle\tilde  N_x^4 \rangle /\langle\tilde  N_x^2 \tilde \varphi_x^2 \rangle &= 9/5&
\langle\tilde  \varphi_x^4 \rangle /\langle\tilde  N_x^2 \tilde \varphi_x^2 \rangle &= 3.
\end{align}

{\bf Regime with only lattice symmetry (low $T$).} We average $\vec N$ over the 6 directions $\vec{\tilde  N} = \sqrt{3} (1,0,0)$, $\vec {\tilde N} = \sqrt{3} (-1,0,0)$ etc. giving  $\langle\tilde  N_x^{2k} \rangle = 3^{k-1}$,
\begin{align}
\langle \tilde N_x^4\rangle/\langle\tilde  \varphi_x^4 \rangle &=1&
\langle \tilde N_x^4 \rangle /\langle \tilde N_x^2 \tilde \varphi_x^2 \rangle &= 3&
\langle \tilde \varphi_x^4 \rangle /\langle \tilde N_x^2 \tilde \varphi_x^2 \rangle &= 3.
\end{align}

\section{Results for smaller \(v_4\)}

We examine the transition at four points on the critical line, for
\begin{equation}
v_4 = 0, \, 0.2, \, 1, \, 10.
\end{equation}
Previous work found an apparent tricritical point near $v_4=0$, with a first order transition for $v_4\lesssim 0$. 

\begin{figure}
\includegraphics[width=0.5\columnwidth]{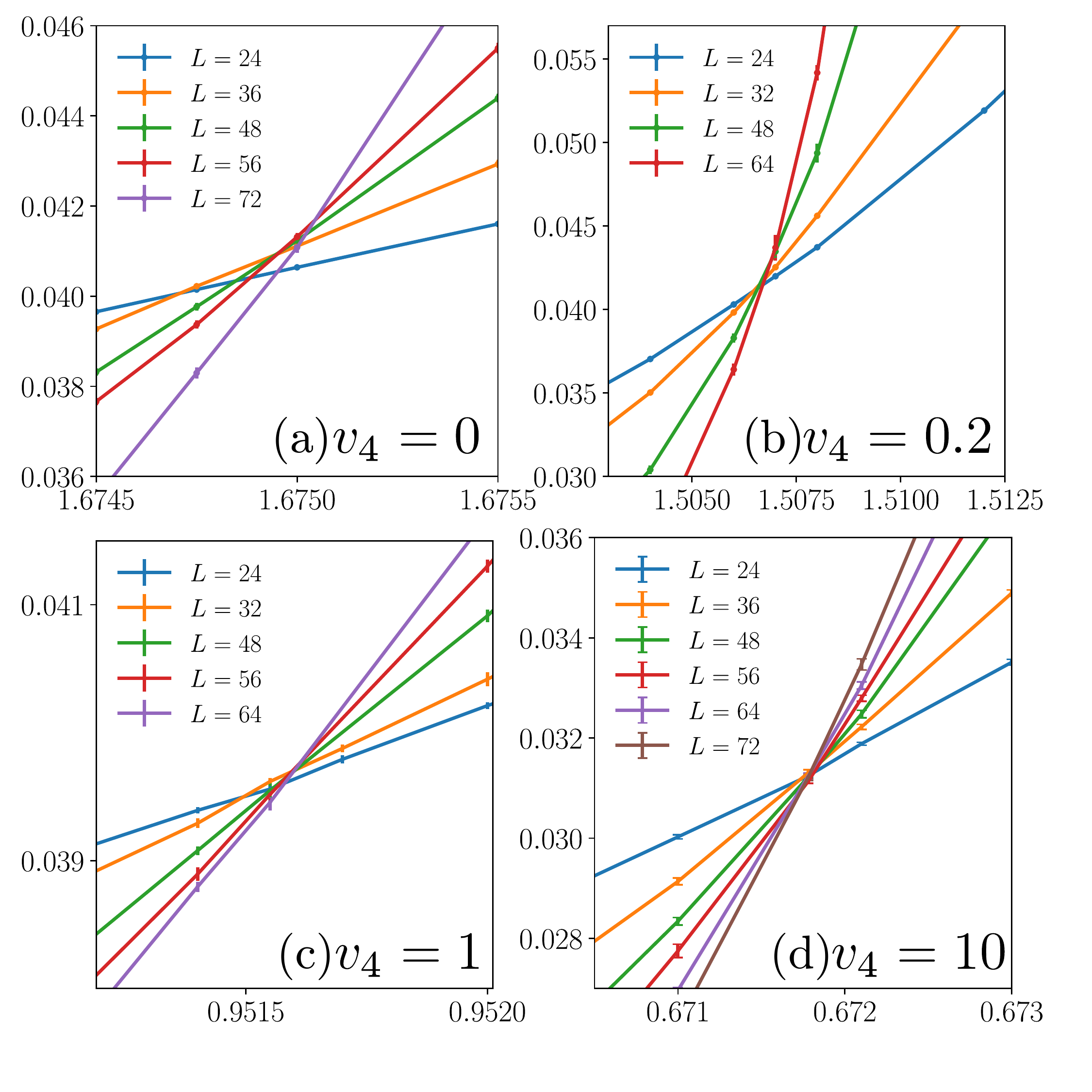}
\caption{
{
The ratio $\langle \varphi_x^2 \rangle/\langle N_x^2 \rangle$ as a function of temperature close to $T_c$, for four values of the frustrating interaction $v_4$. Note that the scales differ between the plots.
}
}
\label{phiSqByNSq}
\end{figure}

\begin{figure}
\includegraphics[width=0.55\columnwidth]{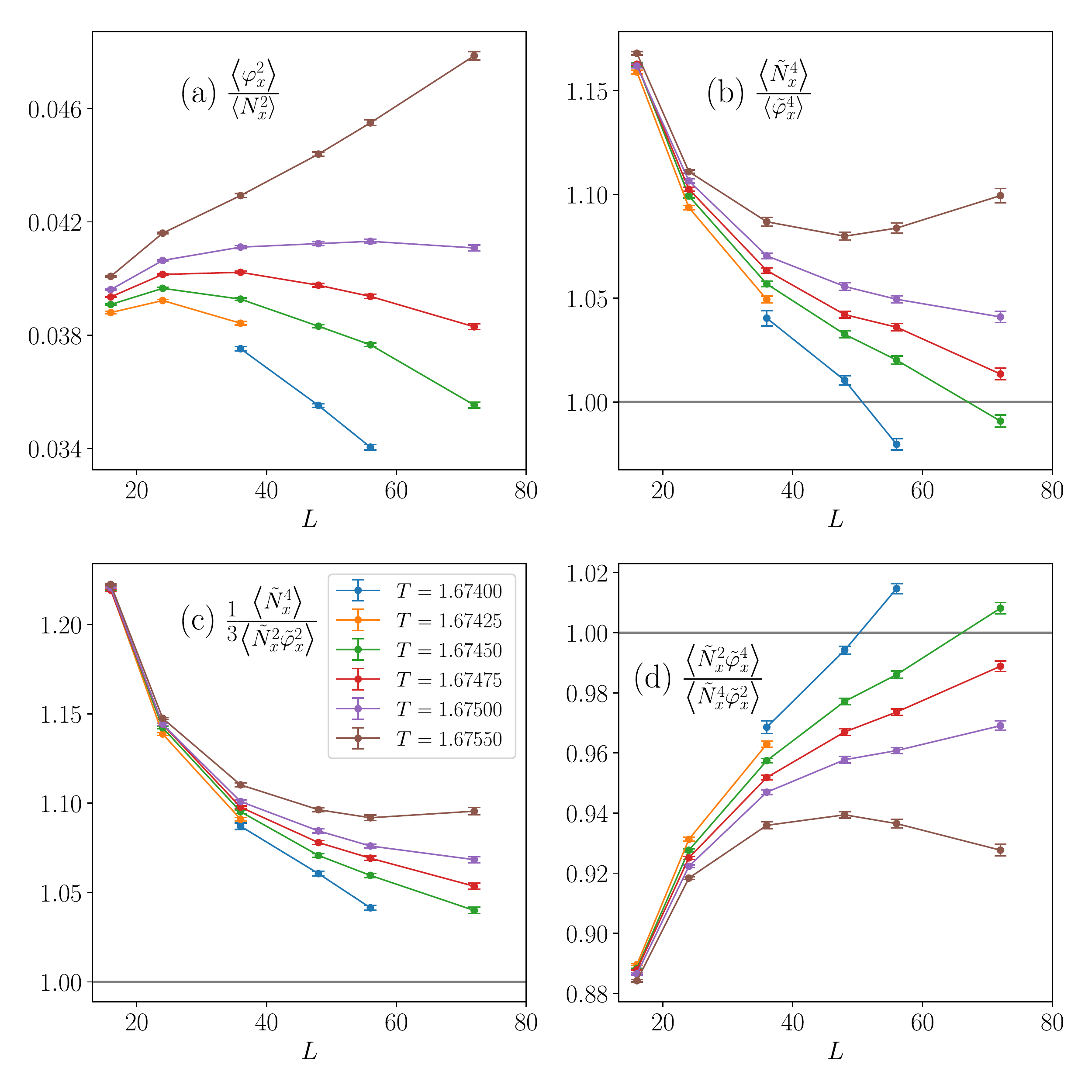}
\caption{
{
Data as a function of $L$ in the vicinity of the apparent tricritical point (for $v_4=0$), for temperatures close to $T_c$.  
(a) $\langle \varphi_x^2 \rangle/\langle N_x^2 \rangle$, which should be $L$-independent at $T_c$ in the presence of $\SO(5)$. (b-c) Moment ratios that should tend to unity, for $T=T_c$, in the presence of $\SO(5)$.
}
}
\label{v4_0_constantTlines}
\end{figure}

\begin{figure}
\includegraphics[width=0.55\columnwidth]{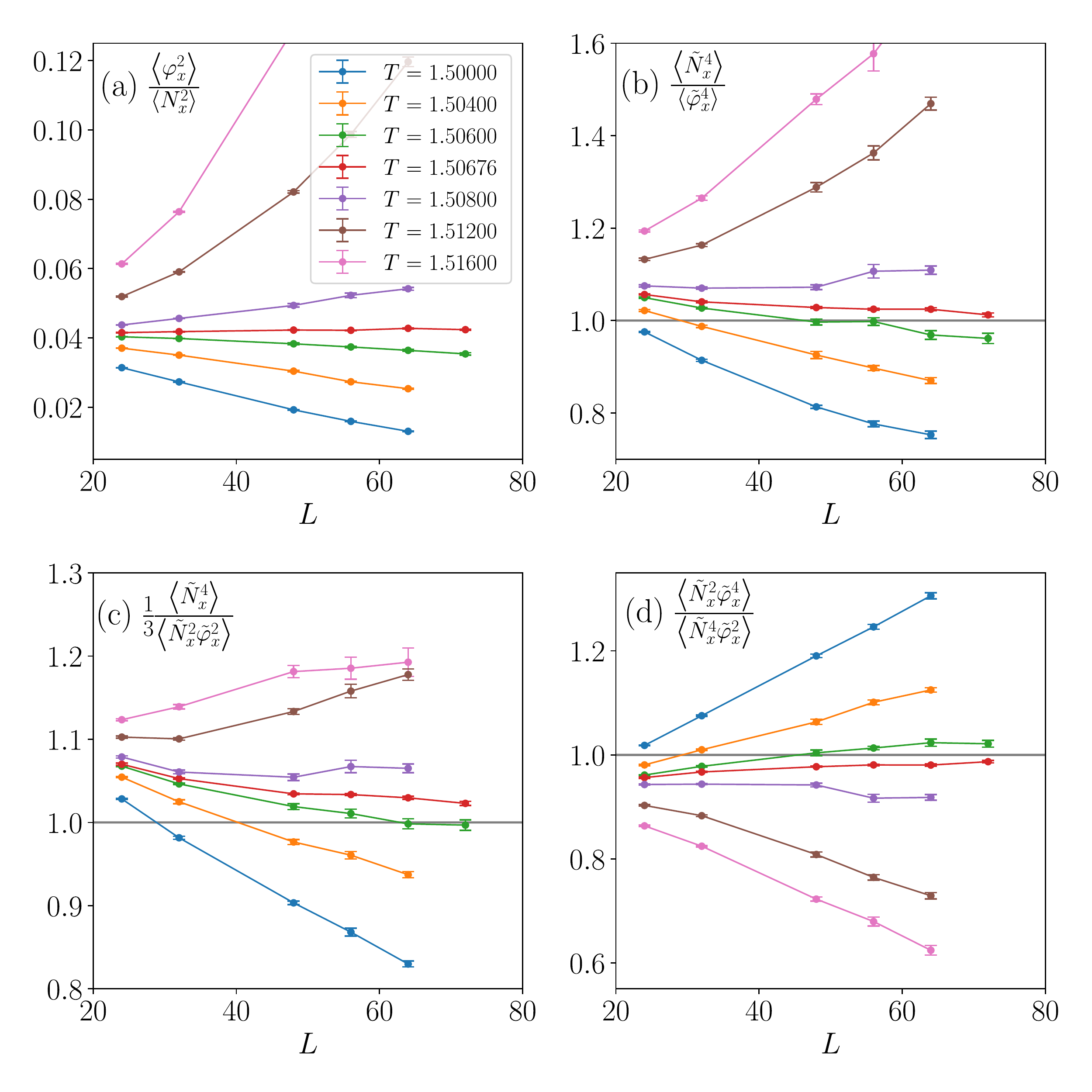}
\caption{
{
Data as a function of $L$ at $v_4=0.2$, for temperatures close to $T_c$.  
(a) $\langle \varphi_x^2 \rangle/\langle N_x^2 \rangle$, which should be $L$-independent at $T_c$ in the presence of $\SO(5)$. (b-c) Moment ratios that should tend to unity, for $T=T_c$, in the presence of $\SO(5)$.
}
}
\label{v4_0p2_constantTlines}
\end{figure}

\begin{figure}
\includegraphics[width=0.55\columnwidth]{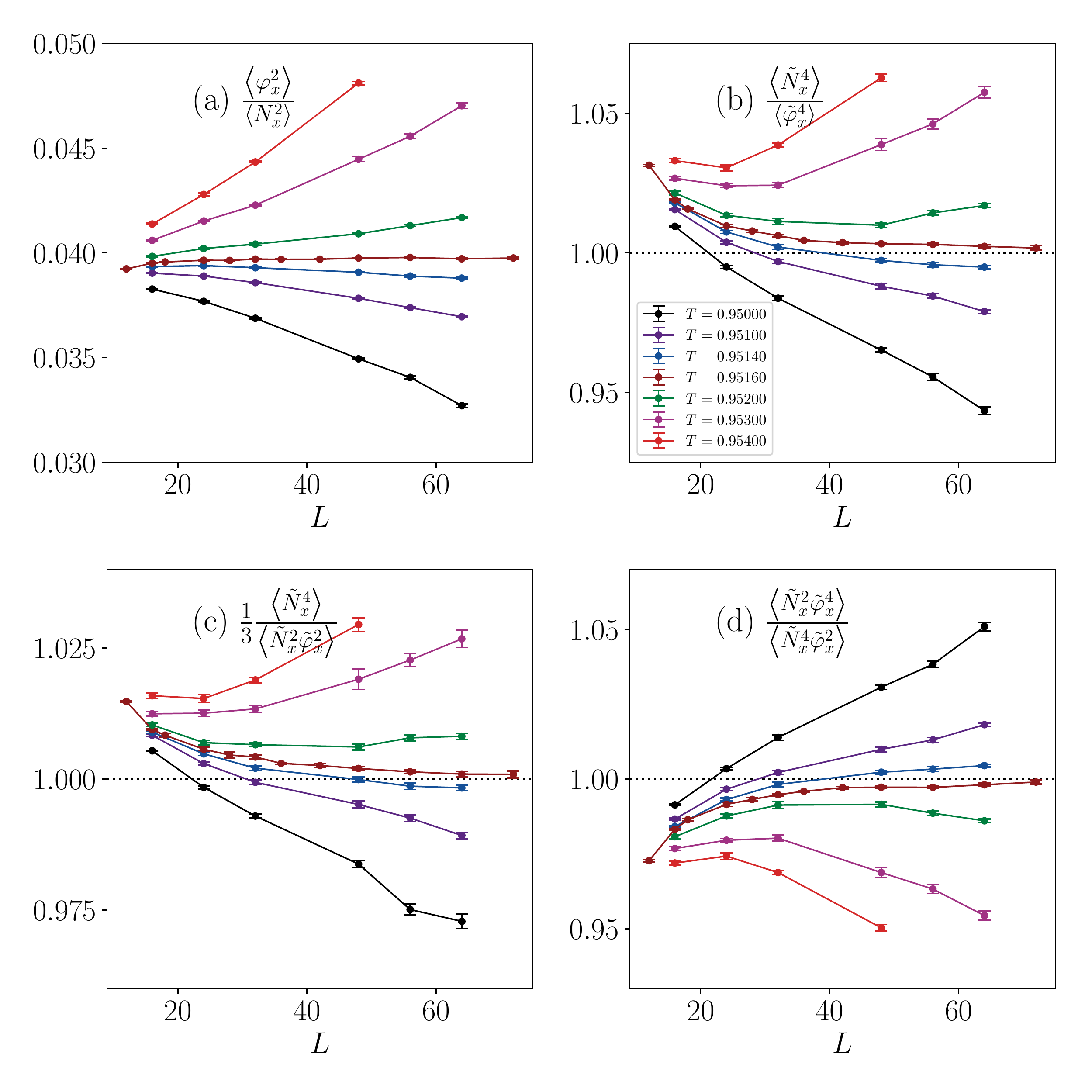}
\caption{
{
Data as a function of $L$ at $v_4=1$, for temperatures close to $T_c$.  
(a) $\langle \varphi_x^2 \rangle/\langle N_x^2 \rangle$, which should be $L$-independent at $T_c$ in the presence of $\SO(5)$. (b-c) Moment ratios that should tend to unity, for $T=T_c$, in the presence of $\SO(5)$.
}
}
\label{v4_1_constantTlines}
\end{figure}

\begin{figure}
\includegraphics[width=0.55\columnwidth]{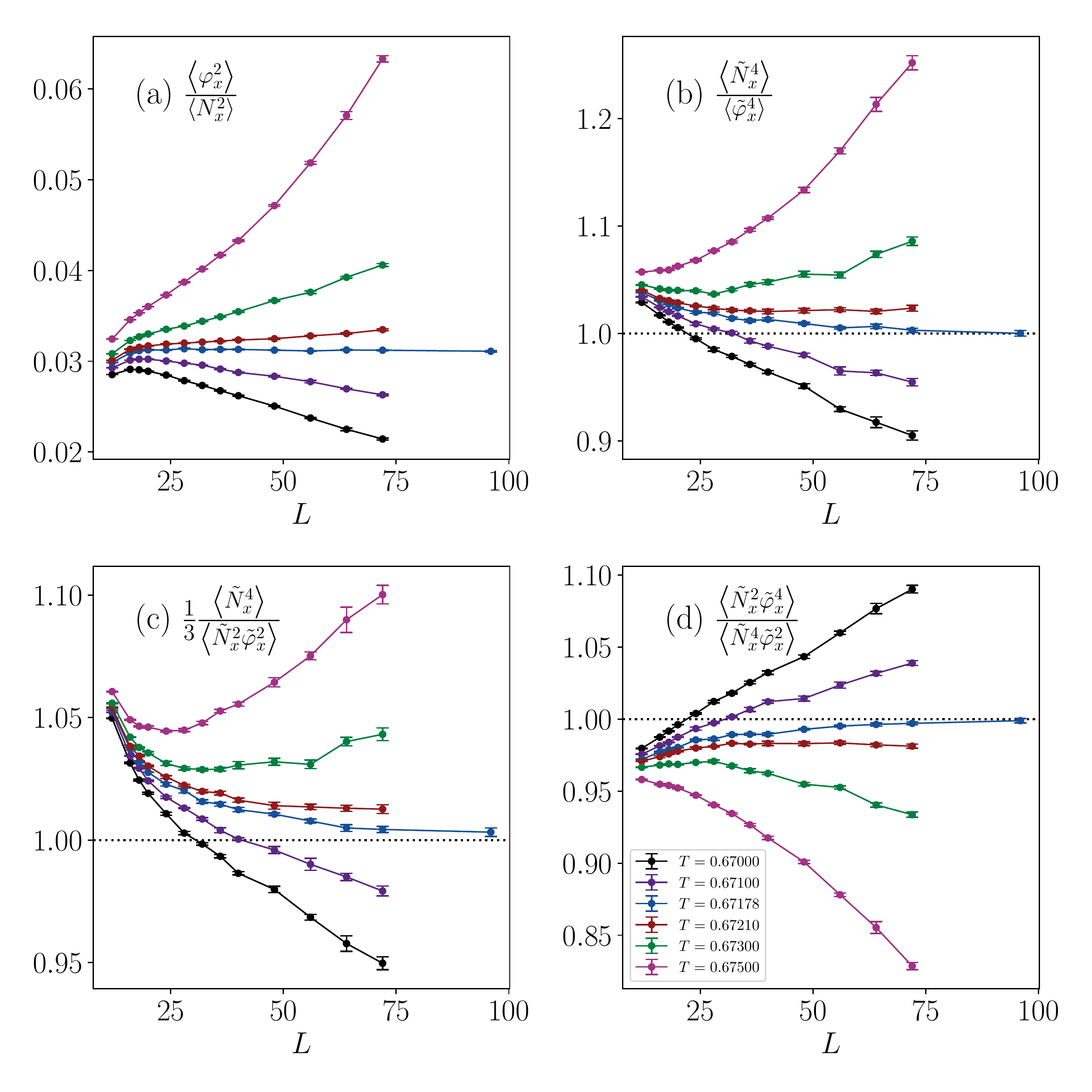}
\caption{
{
Data as a function of $L$ at $v_4=10$, for temperatures close to $T_c$.  
(a) $\langle \varphi_x^2 \rangle/\langle N_x^2 \rangle$, which should be $L$-independent at $T_c$ in the presence of $\SO(5)$. (b-c) Moment ratios that should tend to unity, for $T=T_c$, in the presence of $\SO(5)$.
}
}
\label{v4_10_constantTlines}
\end{figure}

Figure~\ref{phiSqByNSq} shows the crossings of \( \langle \varphi_x^2 \rangle/ \langle N_x^2 \rangle\) as a function of $T$ for these \(v_4\) values.  In Figures~\ref{v4_0_constantTlines}--\ref{v4_10_constantTlines} we show \(\SO(5)\)-sensitive ratios versus \(L\) for several temperatures near the transition. 

The departure of the ratios from the $\SO(5)$-invariant values is much smaller (at the largest sizes) for the  two larger values of $v_4$. One possibility is that $v_4\sim 0$ is not a tricritical point, but rather a sharp crossover from a `quasiuniversal', very weakly first order regime at $v_4\gtrsim 0$, to a strongly first order one at $v_4\lesssim 0$. If so then we would expect the accuracy of $\SO(5)$ to decrease as $v_4$ tends towards $v_4\sim 0$ from above, regardless of how large a system size is explored. Nevertheless it should be noted that for $v_4=0.2$ the ratios are within a few percent of the $\SO(5)$ values (assuming $T_c$ is between $1.50600$ and $1.50676$), and for both $v_4=0.2$ and $v_4=0$ the quality of $\SO(5)$ may improve at larger sizes.

\begin{figure}
\includegraphics[width=0.75\columnwidth]{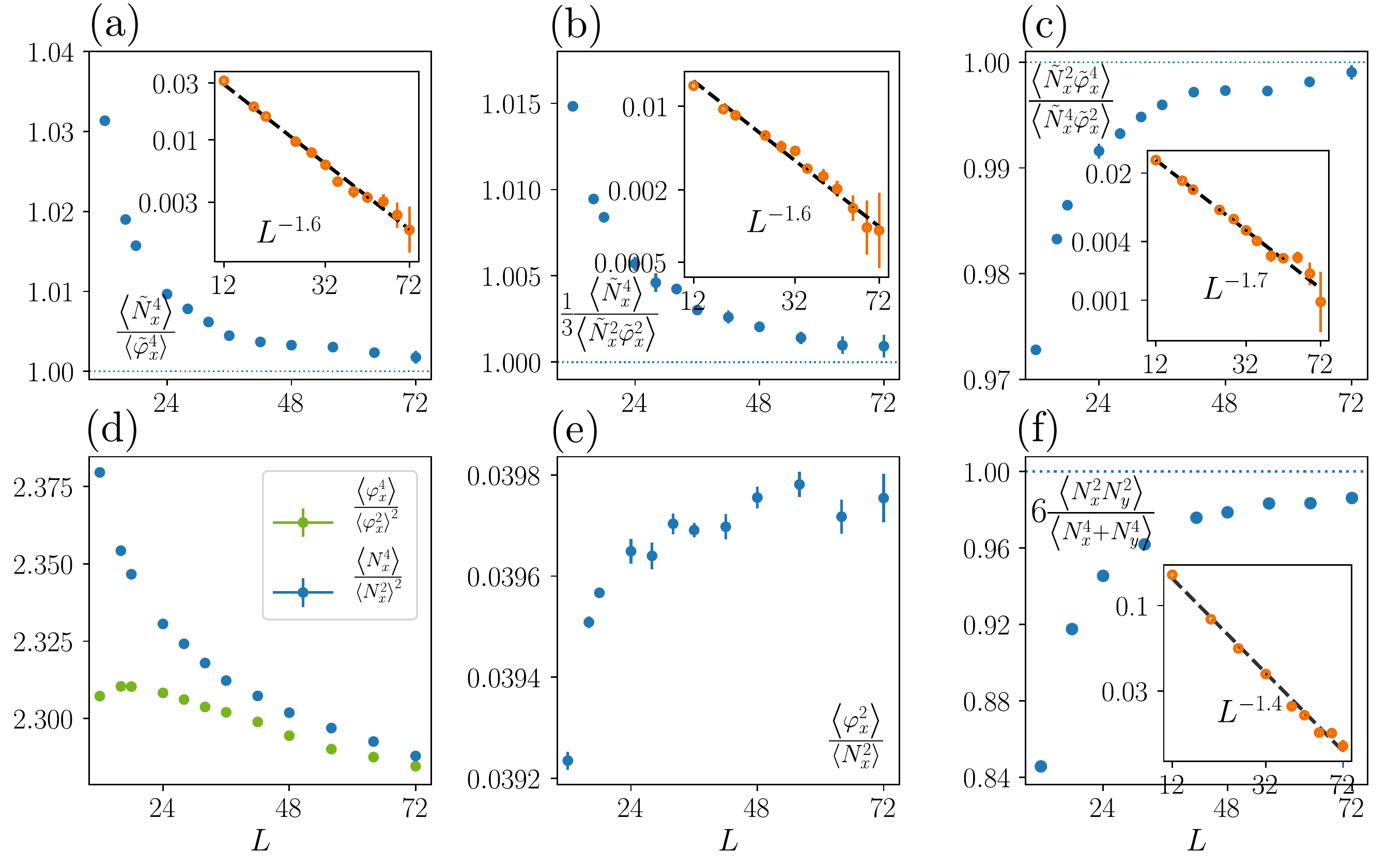}
\caption{{
Data for various moment ratios as a function of $L$, at $v_4=1$ and at the estimated $T_c = 0.95160$. Panels (a-c) and (f) show approximately power-law convergence to the $\SO(5)$-invariant value of unity (this is similar to that shown in the main text $v_4=10$, but with slightly larger effective exponents). 
In panel (d) we see that the variations in the individual Binder cumulants $\langle \varphi_x^4\rangle/\langle \varphi_x^2\rangle^2$ and $\langle N_x^4\rangle/\langle N_x^2\rangle^2$  over these lengthscales are of the same order of magnitude as the variation of the ratio in (a) that is sensitive to $\SO(5)$.}}
\label{v4_1_2}
\end{figure}

Figure~\ref{v4_1_2} examines the case $v_4=1$ in more detail, showing how these same ratios approach unity as a function of system size at the critical temperature. The trends are similar to those for $v_4=10$, and the moments are very close to 1 at large $L$, providing evidence that emergent $\SO(5)$ is a generic property of the transition in the regime where it is continuous/quasicontinuous.

\begin{thebibliography}{99}

\frenchspacing
\raggedright

\newcommand{\journal}[4]{#1 {\bf #2}, #3 (#4)}
\newcommand{\journaltitle}[5]{{\it #1}, \journal{#2}{#3}{#4}{#5}}
\newcommand{\journaltitlenovolume}[4]{{\it #1}, #2 ({\bf #4}), #3}
\newcommand{\book}[3]{{\it #1} (#2, #3)}
\newcommand{\bookeds}[4]{{\it #1}, #2 (#3, #4)}
\newcommand{\pr}{Phys. Rev.}
\newcommand{\prx}{Phys. Rev. X}

\bibitem{Huse2003} D. A. Huse, W. Krauth, R. Moessner, and S. Sondhi, \journaltitle{Coulomb and liquid dimer models in three dimensions}{\prl}{91}{167004}{2003}.
\bibitem{Henley2010} C. L. Henley, \journaltitle{The ``Coulomb phase'' in frustrated systems}{Annu. Rev. Cond. Matt. Phys.}{1}{179}{2010}.
\bibitem{Alet2006} F. Alet, G. Misguich, V. Pasquier, R. Moessner, and J. L. Jacobsen, \journaltitle{Unconventional Continuous Phase Transition in a Three-Dimensional Dimer Model}{\prl}{97}{030403}{2006}.
\bibitem{PowellChalker} S. Powell and J. T. Chalker, \journaltitle{\(\mathrm{SU}(2)\)-Invariant Continuum Theory for an Unconventional Phase Transition in a Three-Dimensional Classical Dimer Model}{\prl}{101}{155702}{2008}; \journaltitle{Classical to quantum mapping for an unconventional phase transition in a three-dimensional classical dimer model}{\prb}{80}{134413}{2009}.
\bibitem{Charrier2008} D. Charrier, F. Alet, and P. Pujol, \journaltitle{Gauge Theory Picture of an Ordering Transition in a Dimer Model}{\prl}{101}{167205}{2008}.
\bibitem{Chen2009} G. Chen, J. Gukelberger, S. Trebst, F. Alet, and L. Balents, \journaltitle{Coulomb gas transitions in three-dimensional classical dimer models}{\prb}{80}{045112}{2009}.
\bibitem{Senthil2004} T. Senthil, A. Vishwanath, L. Balents, S. Sachdev, and M. P. A. Fisher, \journaltitle{Deconfined Quantum Critical Points}{Science}{303}{1490}{2004}.
\bibitem{Senthil2004a} T. Senthil, L. Balents, S. Sachdev, A. Vishwanath, and M. P. A. Fisher, \journaltitle{Quantum criticality beyond the Landau-Ginzburg-Wilson paradigm}{\prb}{70}{144407}{2004}.
\bibitem{sandvikPRL2007} A. W. Sandvik, \journaltitle{Evidence for Deconfined Quantum Criticality in a Two-Dimensional Heisenberg Model with Four-Spin Interactions}{\prl}{98}{227202}{2007}.
\bibitem{MelkoKaul} R. G. Melko and R. K. Kaul, \journaltitle{Scaling in the fan of an unconventional quantum critical point}{\prl}{100}{017203}{2008}, 
\bibitem{Kuklov2008} A. B. Kuklov, M. Matsumoto, N. V. Prokof'ev, B. V. Svistunov, and M. Troyer, 
\journaltitle{Deconfined criticlaity: generic first-order transition in the $SU(2)$ symmetry case}{\prl}{101}{050405}{2008}
\bibitem{Jiang2008} F.-J. Jiang, M. Nyfeler, S. Chandrasekharan, and U.-J. Wiese, \journaltitle{From an antiferromagnet to a valence bond solid: evidence for a first-order phase transition}{Journal of Statistical Mechanics: Theory and Experiment}{2008}{P02009}{2008}
\bibitem{LouSandvikKawashima} J. Lou, A. W. Sandvik, and N. Kawashima,
\journaltitle{Antiferromagnetic to valence-bond-solid transitions in two-dimensional $SU(N)$ Heisenberg models with multispin interactions}{\prb}{80}{180414}{2009}
\bibitem{BanerjeeDamleAlet} A. Banerjee, K. Damle, and F. Alet, 
\journaltitle{Impurity spin texture at a deconfined quantum critical point}{\prb}{82}{155139}{2010}
\bibitem{Sandvik2010} A. W. Sandvik, \journaltitle{Continuous Quantum Phase Transition between an Antiferromagnet and a Valence-Bond Solid in Two Dimensions: Evidence for Logarithmic Corrections to Scaling}{\prl}{104}{177201}{2010}.
\bibitem{Harada2013} K. Harada, T. Suzuki, T. Okubo, H. Matsuo, J. Lou, H. Watanabe, S. Todo, and N. Kawashima,
\journaltitle{Possibility of deconfined criticality in $SU(N)$ Heisenberg models at small $N$}{\prb}{88}{220408}{2013}
\bibitem{Chen2013} K. Chen, Y. Huang, Y. Deng, A. B. Kuklov, N. V. Prokof'ev, and B. V. Svistunov, \journaltitle{Deconfined Criticality Flow in the Heisenberg Model with Ring-Exchange Interactions}{\prl}{110}{185701}{2013}.
\bibitem{ShaoGuoSandvik} H. Shao, W. Guo, and A. W. Sandvik, \journaltitle{Quantum criticality with two length scales}{Science}{352}{213}{2016}
\bibitem{Nahum2015a} A. Nahum, P. Serna, J. T. Chalker, M. Ortu{\~n}o, and A. M. Somoza, \journaltitle{Deconfined Quantum Criticality, Scaling Violations, and Classical Loop Models}{\prx}{5}{041048}{2015}.
\bibitem{Nahum2015} A. Nahum, P. Serna, J. T. Chalker, M. Ortu{\~n}o, and A. M. Somoza, \journaltitle{Emergent \(\SO(5)\) symmetry at the N\'eel to valence-bond-solid transition}{\prl}{115}{267203}{2015}.
\bibitem{Tanaka2005} A. Tanaka and X. Hu, \journaltitle{Many-Body Spin Berry Phases Emerging from the \(\pi\)-Flux State: Competition between Antiferromagnetism and the Valence-Bond-Solid State}{\prl}{95}{036402}{2005}.
\bibitem{Senthil2006} T. Senthil and M. P. A. Fisher, \journaltitle{Competing orders, nonlinear sigma models, and topological terms in quantum magnets}{\prb}{74}{064405}{2006}.
\bibitem{Wang2017} C. Wang, A. Nahum, M. A. Metlitski, C. Xu, and T. Senthil, \journaltitle{Deconfined Quantum Critical Points: Symmetries and Dualities}{\prx}{7}{031051}{2017}.
\bibitem{Suwa2016} H. Suwa, A. Sen, and A. W. Sandvik, \journaltitle{Level spectroscopy in a two-dimensional quantum magnet: Linearly dispersing spinons at the deconfined quantum critical point}{\prb}{94}{144416}{2016}.
\bibitem{Bartosch} L. Bartosch, \journaltitle{Corrections to scaling in the critical theory of deconfined criticality}{\prb}{88}{195140}{2013}
\bibitem{SimmonsDuffin}  D. Simmons-Duffin, unpublished.
\bibitem{Nakayama} Y. Nakayama and T. Ohtsuki, \journaltitle{Necessary condition for emergent symmetry from the conformal bootstrap}{\prl}{117}{131601}{2016}
\bibitem{Charrier2010} D. Charrier and F. Alet, \journaltitle{Phase diagram of an extended classical dimer model}{\prb}{82}{014429}{2010}.
\bibitem{MotrunichVishwanath} O. I. Motrunich and A. Vishwanath, \journaltitle{Emergent photons and transitions in the O(3) sigma model with hedgehog suppression}{\prb}{70}{075104}{2004}
\bibitem{ProkofevSvistunov} N. Prokof'ev and B. Svistunov, \journaltitle{Worm algorithms for classical statistical models}{\prl}{87}{160601}{2001}
\bibitem{Sandvik2006} A. W. Sandvik and R. Moessner, \journaltitle{Correlations and confinement in nonplanar two-dimensional dimer models}{\prb}{73}{144504}{2006}.
\bibitem{Sreejith2014} G. J. Sreejith and S. Powell, \journaltitle{Critical behavior in the cubic dimer model at nonzero monomer density}{\prb}{89}{014404}{2014}.
\bibitem{Misguich2008} G. Misguich, V. Pasquier, and F. Alet, \journaltitle{Correlations and order parameter at a Coulomb-crystal phase transition in a three-dimensional dimer model}{\prb}{78}{100402(R)}{2008}.
\bibitem{Meng2017} Y. Q. Qin, Y.-Y. He, Y.-Z. You, Z.-Y. Lu, A. Sen, A. W. Sandvik, C. Xu, and Z. Y. Meng, \journaltitle{Duality between the deconfined quantum critical point and the bosonic topological transition}{\prx}{7}{031052}{2017}
\bibitem{ZhangHe2018} X.-F. Zhang, Y.-C. He, S. Eggert, R. Moessner, and F. Pollmann, \journaltitle{Continuous easy-plane deconfined phase transition on the kagome lattice}{\prl}{120}{115702}{2018}

\bibitem{anomalyfootnote} More formally, emergent \(\SO(5)\) is consistent with the anomaly structure of the \NCCP\ model \cite{Wang2017}.
\bibitem{scalingfootnote}{{These include constraints on scaling dimensions from conformal bootstrap \cite{SimmonsDuffin,Nakayama} and violations of finite-size scaling in Monte Carlo simulations. See \cite{Wang2017} for a discussion.}}
\bibitem{symmetriesfootnote}{{The spatial rotation and reflection symmetries of the cubic lattice (with the origin at a site) also act as rotations and reflections on \(\Nv\). Translation by a lattice vector changes the sign of one component of \(\Nv\), and also changes the sign of $\varphi_y$, since it exchanges monopoles and antimonopoles.}}
\bibitem{jointdistributionfootnote}{{
In the dimer model one can directly measure the joint probability distribution of three components $(N_x, N_y, N_z)$ of $\Phi$, but not of all five. 
In the loop model it is also possible to measure the joint probability distribution for at most three order parameter components simultaneously: in that case, two components of the VBS and one component of the N\'eel order parameter.
The  imaginary (topological) term in the action for the sigma model \cite{Tanaka2005,Senthil2006}
suggests that a positive joint probability distribution for all five (or even for four) components does not exist even in the continuum.}}
\bibitem{pixelfootnote} In fact \(\lvert N_z \rvert < 0.00625\) (the half-width of one pixel in these plots).
\bibitem{operatorsfootnote}{{At the level of symmetry, $X_{abcd}\sim \Phi_a\Phi_b\Phi_c\Phi_d - (\ldots)$, with subtractions to make $X$ traceless: $\sum_{a=1}^5 X_{aacd}=0$}}
\bibitem{pseudocriticalfootnote}{{In this scenario $\mathrm{SO}(5)$ symmetry is likely to persist beyond the `pseudocritical' regime of lengthscales, into a broad regime where the system resembles a sigma model with full $\mathrm{SO}(5)$ symmetry in its \textit{ordered}, or Goldstone, phase \cite{Wang2017}.}}


\end{thebibliography}
\end{document}